\documentclass[aps,prb,twocolumn]{revtex4-1}
\usepackage{amsmath}
\usepackage{amssymb}
\usepackage{natbib}
\usepackage{physics}
\usepackage{hyperref}
\usepackage{graphicx}
\usepackage{color}

\begin{document}
	\title{Phase-dependent Spin Polarization of Cooper Pairs in Magnetic Josephson Junctions}
	\author{Samme M. Dahir, Anatoly F. Volkov, and Ilya M. Eremin}
	\affiliation{Institut f\"ur Theoretische Physik III,
		Ruhr-Universit\"{a}t Bochum, D-44780 Bochum, Germany}
	\date{\today }

	\begin{abstract}
		Superconductor-Ferromagnet hybrid structures (SF) have attracted much interest in the last decades, due to a variety of interesting phenomena predicted and observed in these structures. One of them is the so-called inverse proximity effect. It is described by a spin polarization of Cooper pairs, which occurs not only in the ferromagnet (F), but also in the superconductor (S) yielding a finite magnetic moment \(M_{\text{S}}\) inside the superconductor. This effect has been predicted and experimentally studied. However, interpretation of the experimental data is mostly ambiguous. Here, we study theoretically the impact of the spin polarized Cooper pairs on the Josephson effect in an SFS junction. We show that the induced magnetic moment \(M_{\text{S}}\) does depend on the phase difference \(\varphi\) and therefore, will  oscillate in time with the Josephson frequency \(2eV/\hbar\) if the current exceeds a critical value. Most importantly, the spin polarization in the superconductor  causes a significant change in the Fraunhofer pattern, which can be easily accessed experimentally.
	\end{abstract}
	
	\maketitle
	
	\date{\today}
	In the past decades, the interest in studying proximity effects in superconductor(S)-ferromagnet(F) heterostructures including the magnetic Josephson junctions has steadily increased\cite{golubov_current-phase_2004,buzdin_proximity_2005,bergeret_odd_2005,eschrig_spin-polarized_2015,linder_superconducting_2015,balatsky_odd_2017}. A number of interesting phenomena were originally predicted and experimentally verified  in various systems. Perhaps the most known among them is the sign change of the Josephson current \(I_{J}\) in SFS junctions (so-called $0-\pi$ shift), predicted  theoretically in Refs. \cite{buzdin1982critical,buzdin1991josephson} and consequently observed in several experiments\cite{ryazanov_coupling_2001,kontos_josephson_2002,sellier_temperature-induced_2003,oboznov_thickness_2006}. The sign change of the critical current \(I_{c}\) is related to the spatial oscillations of the condensate function, induced in the F layer by the proximity effect (PE). Another exciting effect is related to the prediction\cite{bergeret_enhancement_2001,bergeret_odd_2005,eschrig_spin-polarized_2015,linder_superconducting_2015} and observation\cite{keizer_spin_2006,sosnin_superconducting_2006,khaire_observation_2010,anwar_long-range_2010,anwar_long_2012,sprungmann_evidence_2010,robinson_enhanced_2010,blamire_interface_2014,massarotti_electrodynamics_2018,kalenkov_triplet_2011,klose_optimization_2012,martinez_amplitude_2016,niedzielski_spin-valve_2018,caruso_tuning_2019} of a long-range triplet component of the condensate wave function in the SF bilayer structures in the presence of the inhomogeneous magnetization \(M_{\text{F}}\) in the ferromagnet F. Note, that the triplet even-parity component, which due to the Pauli principle has to be odd in frequency, arises in SF systems with both homogeneous and inhomogeneous magnetizations. However, in the case of a homogeneous magnetization, the total spin of the triplet Cooper pairs has zero projection on the direction of magnetization, \(\vb{M}_{\text{F}}\), and thus the condensate wave function penetrates the ferromagnet over a short distance \(\xi_{\text{s}-r}=\xi_{\text{F}}\) only, where in the diffusive case \(\xi_{\text{F}}\cong\sqrt{D_{\text{F}}/J}\). Here, \(D_{\text{F}}\) and \(J\) are the diffusion coefficient and the exchange field, respectively. At the same time, a non-homogeneous magnetization leads to a finite projection of the total spin of the triplet Cooper pair along \(\vb{M}_{\text{F}}\). As a result,  the condensate wave function penetrates over much longer distances \(\xi_{l-r}\cong\sqrt{D_{\text{F}}/2\pi T}\)\cite{bergeret_enhancement_2001,bergeret_odd_2005,eschrig_spin-polarized_2015,linder_superconducting_2015}. The presence of the long-range triplet component was confirmed in various experiments\cite{keizer_spin_2006,sosnin_superconducting_2006,khaire_observation_2010,anwar_long-range_2010,anwar_long_2012,sprungmann_evidence_2010,robinson_enhanced_2010,blamire_interface_2014,massarotti_electrodynamics_2018,kalenkov_triplet_2011,klose_optimization_2012,martinez_amplitude_2016,niedzielski_spin-valve_2018,caruso_tuning_2019}. For example, 
	in S\(\text{F}_{\text{ml}}\)S Josephson junctions with a multilayered ferromagnet \(\text{F}_{\text{ml}}\) it was found \cite{khaire_observation_2010,klose_optimization_2012,martinez_amplitude_2016} that the Josephson current \(I_{J}\) is only present if the magnetization vectors of the different F layers were non-collinear, while for collinear \(\vb{M}_{\text{F}}\) magnetization orientation, the Josephson current was negligibly small. 
	
	In addition to the direct proximity effect in SF heterostructures, describing the penetration of Cooper pairs from S to F, there exists also the so-called inverse proximity effect.\cite{krivoruchko_inhomogeneous_2002,bergeret_manifestation_2003,bergeret_spin_2004,bergeret_induced_2004,lofwander_interplay_2005,Faure_theory_2007} The latter is characterized by an induced magnetic moment \(M_{\text{S}}\) in the superconductor S. This induced magnetization \(M_{\text{S}}(z)\) decays inside the superconductor on a distance \(z\) of the order of the superconducting coherence length \(\xi_{S}\sim\sqrt{d_{\text{S}}/\Delta}\), where $2d_{\text{S}}$ is a superconductor thickness. In the diffusive superconductors, \textit{i.e.} when the
	concentration of non-magnetic impurities in a system is
	sufficiently large, the vector \(\vb{M}_{\text{S}}\) is aligned in the direction opposite to \(\vb{M}_{\text{F}}\). Under appropriate conditions and at low temperatures, a full spin screening may take place in these superconductors, \textit{i.e.}, the total magnetization \(\mathcal{M}_{\text{S}}\) induced in the superconductor S may be equal to \(-2d_{\text{F}}M_{\text{F}}\), where \(2d_{\text{F}}\) is the thickness of the F layer. At the same time, in the ballistic case the magnetization \(M_{\text{S}}(z)\) was found to oscillate with \(z\) \cite{bergeret_inverse_2005,kharitonov_oscillations_2006}.
	There were experimental attempts  to observe the induced magnetization in S \cite{salikhov_experimental_2009,xia_inverse_2009} and although a magnetic field in the S film was detected in several experiments, the interpretation of these results was still ambiguous due to signal to noise ratio and also by the fact that the induced magnetization (spin polarization) was masked by a magnetic field \(H_{orb}\) created by spontaneous Meissner currents (orbital effects). Theoretically, the interplay between spin polarization and orbital effects has been analyzed in Ref. \cite{bergeret_spin_2004} and in more detail more recently\cite{volkov_spin_2019}. In addition the calculation of \(H_{orb}\) without taking into account the induced magnetization \(M_{\text{S}}\) was carried out in Ref. \cite{mironov_electromagnetic_2018,devizorova_electromagnetic_2019}. Further efforts are needed to unambiguously prove the existence of the inverse proximity effect. One must therefore look for clear manifestation of the induced magnetization which are experimentally accessible.\\
	
\begin{figure}[tbp]
		\includegraphics[width=\columnwidth]{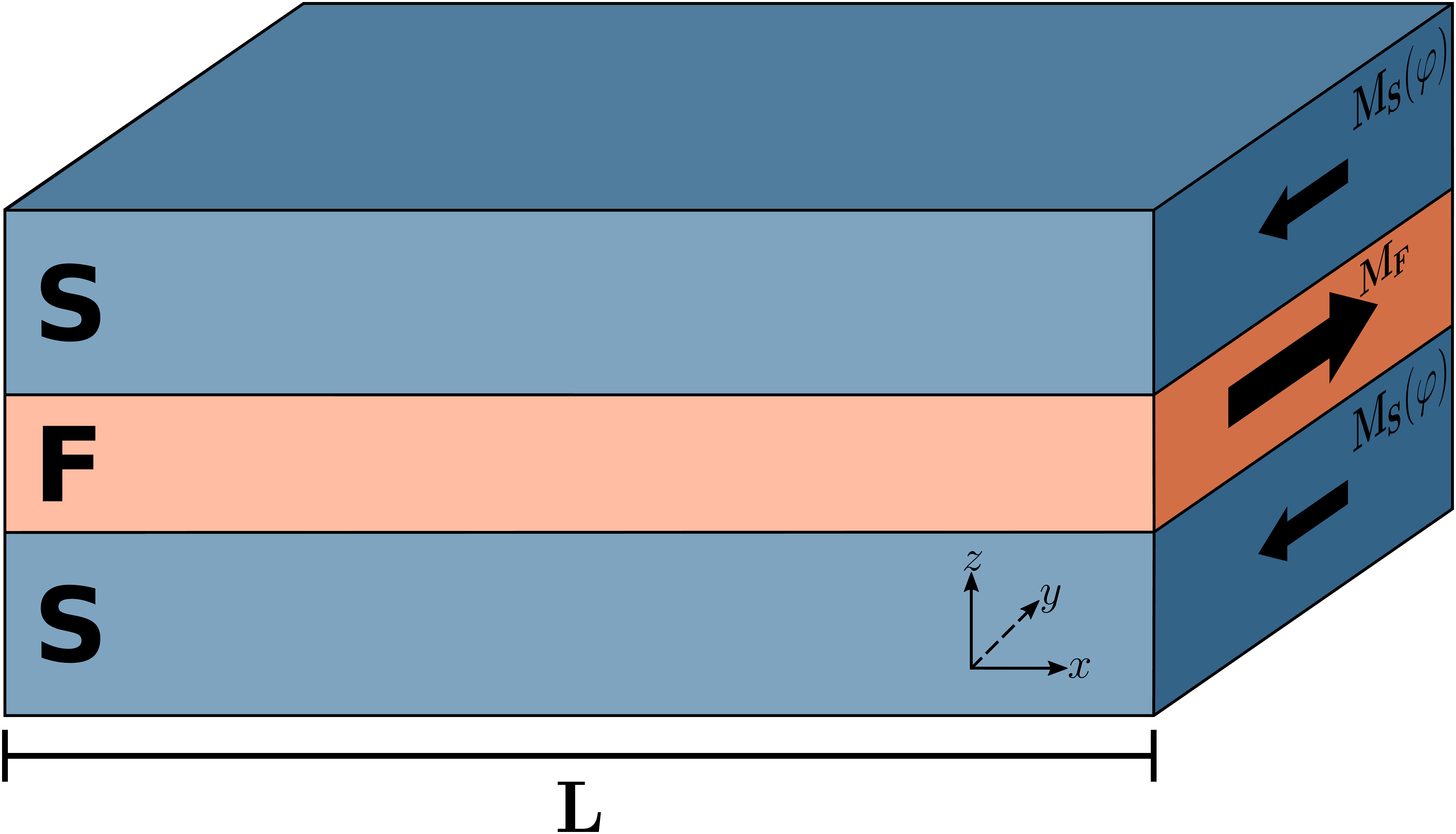}
		\caption{(Color online) Schematic representation of the considered SFS
			Josephson junction. The thickness of the F film is $2d_{\text{F}}$. The S films are
			assumed to be much thicker than the London penetration depth, \textit{i.e.} \(2d_{\text{S}}\gg\lambda_{\text{S}}\).\label{fig:1 Scheme}}
	\end{figure}

	In the present work, we study the mutual influence of the spin polarization and the Josephson effect in SFS junctions as illustrated in Fig.\ref{fig:1 Scheme}. In particular, we show that the magnetic moment \(M_{\text{S}}\) in the superconductor due to spin polarization does depend on the phase difference \(\varphi\) such that, if the bias current \(I\) exceeds the critical current \(I_{c}\), the magnetic moment \(M_{\text{S}}(\varphi(t))\) oscillates with the Josephson frequency \(\omega_{J}=2eV/\hbar\). Most importantly, the spin polarization in the S films affects the prominent Fraunhofer pattern for the Josephson effect, which can be observed experimentally using sensitive Josephson interferometry. Furthermore, we analyze the situations for high and  low interface transparencies.\\
	
	This work is structured as follows: In section I we investigate the induced magnetization in the S for the case of a weak and a strong PE. The expressions found here are the basis for our study in section II, where we determine the change of magnetostatic quantities in SFS junctions due to the phase dependent contribution of the induced magnetization. In section III, we consider the effect of the spin polarization in the S on the standard Fraunhofer pattern in a SFS junctions. We conclude our work in section IV.
	
	\section{Induced Magnetization in the Superconductor}
	
	In the following we derive a formula for the magnetization \(M_{\text{S}}\) in the SFS junction induced in the S regions. The derivation is similar to that presented in
	Refs. \cite{bergeret_spin_2004,volkov_spin_2019}. Detailed calculations for the strong and weak proximity effect can be found in the Appendix A and B.

\begin{figure}[tbp]
		\includegraphics[width=\columnwidth]{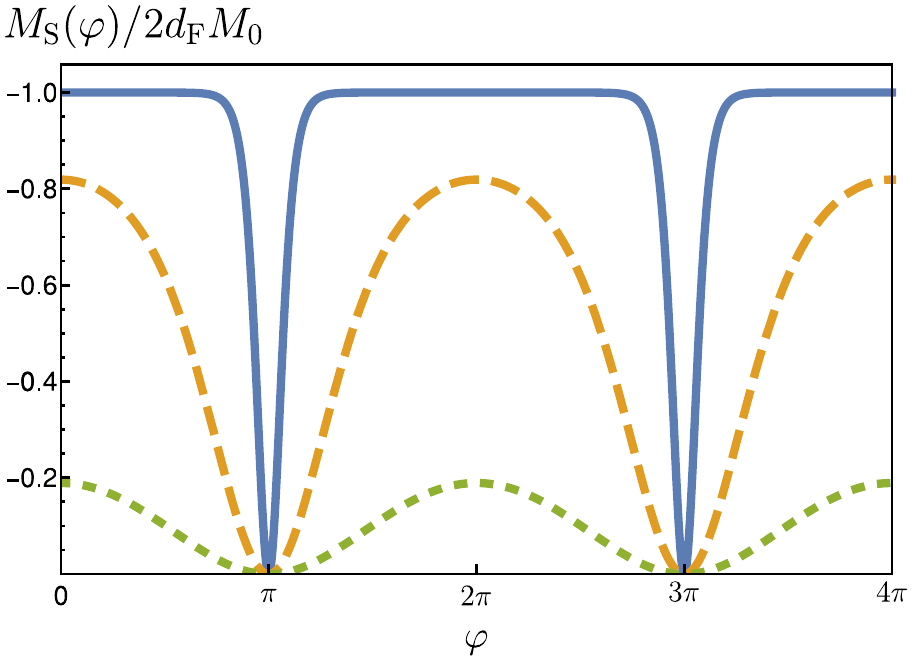}
		\caption{(Color online) The phase dependence of the total induced magnetization \(\mathcal{M}_{\text{S}}(\varphi)\) for the strong proximity effect in the S film normalized by \(2d_{\text{F}}M_{0}\) for various temperatures \(\tilde{T}=T/T_{c}\) with \(\tilde{T}=0.1\) (solid blue curve), \(\tilde{T}=0.5\) (orange dashed curve) and \(\tilde{T}=0.9\) (green dotted curve) \label{fig:2 Phase dependence}}
	\end{figure}

	In the case of SFS Josephson junctions, one has to take into account the phase difference \(\varphi\) which affects the energy spectrum and the density of states (DOS) in the F film. We assume a dirty limit which is
    described by the Usadel equations for the Green's functions  in the superconductor, $\check{g}_{\text{S}}$, and in the ferromagnet, $\check{g}_{\text{F}}$, respectively 
	\begin{widetext}
	\begin{equation}
	-D_{\text{S}}\partial_{z}\qty(\check{g}_{\text{S}}\partial_{z}\check{g}_{\text{S}})+\omega\qty[\check{X}_{30},\check{g}_{\text{S}}]+\Delta \qty[\check{X}_{10}\cos(\varphi/2)\pm\check{X}_{20}\sin(\varphi/2),\check{g}_{\text{S}}]=0, \quad \pm\text{S films}\label{Appendix Strong PE: Usadel S}
	\end{equation}
	\begin{equation}
	-D_{\text{F}}\partial_{z}\qty(\check{g}_{\text{F}}\partial\check{g}_{\text{F}})+\omega\qty[\check{X}_{30},\check{g}_{\text{F}}]+iJ\qty[\check{X}_{33},\check{g}_{\text{F}}]=0,\quad \text{F film}\label{Appendix Strong PE: Usadel F}
	\end{equation}
	where $\Delta$ is the superconducting gap and \(\omega\) is the frequency. Here, \(\check{X}_{ij}\) is defined as \(\check{X}_{ij}=\hat{\tau}_{i}\otimes\hat{\sigma}_{j}\) with the Pauli matrices \(\hat{\tau}\) and \(\hat{\sigma}\) operating in particle-hole (Nambu-Gor'kov) and in spin space.  The \(\pm\)S films denote the superconducting right (left) electrodes.
	Eqs.(\ref{Appendix Strong PE: Usadel S}-\ref{Appendix Strong PE: Usadel F}) are complemented by the normalization relation \(\check{g}_{\text{S}(\text{F})}\cdot\check{g}_{\text{S}(\text{F})}=\check{1}\) and the boundary conditions \cite{Kurpianov1988}
	\begin{align}
	\check{g}_{\text{F}}\partial_{z}\check{g}_{\text{F}}= \kappa_{b\text{F}}\qty[\check{g}_{\text{F}},\check{g}_{\text{S}}] & \qc\quad \check{g}_{\text{S}}\partial_{z}\check{g}_{\text{S}}= \kappa_{b\text{S}}\qty[\check{g}_{\text{S}},\check{g}_{\text{F}}]\label{Appendix Strong PE: Boundary Cond.}
	\end{align}
	where \(\kappa _{b\text{S}(\text{F})}=\varrho_{\text{S}(\text{F})}/R_{b\square}\) with $\varrho_{\text{S}(\text{F})}$ being the resistivity of the S(F) film in the normal state and $R_{b\square}$ is a barrier resistance per unit square.
	
	As in Refs. \onlinecite{bergeret_spin_2004,volkov_spin_2019}, we assume that the Green's functions \(\check{g}_{\text{S}}\) in the S films are only weakly affected by the proximity effect. Thus,  they can be written as
	\begin{equation}
	\check{g}_{\text{S}}^{(0)}=G_{\text{S}}\check{X}_{30}+F_{\text{S}}\qty{\cos(\varphi/2)\check{X}_{10}\pm\sin(\varphi/2)\check{X}_{20}} \label{Appendix Strong PE: BCS Sol}
	\end{equation}
	where \(G_{\text{S}}=\omega/\zeta_{\omega}\) and \(F_{\text{S}}=\Delta/\zeta_{\omega}\) are the normal and anomalous components of the Green's function with \(\zeta_{\omega}=\sqrt{\omega^2+\Delta^{2}}\).	Assuming \(d_{\text{F}}\ll\xi_{\text{F}}=\sqrt{D_{\text{F}}/J}\), Eq.(\ref{Appendix Strong PE: Usadel F}) can be integrated over \(z\) to obtain
	\begin{equation}
	\tilde{\omega}_{\text{F}}\qty[\check{X}_{30},\check{g}_{\text{F}}]+\tilde{\Delta}_{\text{F}}\qty[\check{X}_{10},\check{g}_{\text{F}}]+iJ\qty[\check{X}_{33},\check{g}_{\text{F}}]=0\qc \text{F film} \label{Appendix Strong PE: Integrated Usadel}
	\end{equation}
	where \(\tilde{\omega}_{\text{F}}=(\omega+\epsilon_{b\text{F}}G_{\text{S}})\), \(\tilde{\Delta}_{\text{F}}(\varphi)=\epsilon_{b\text{F}}F_{\text{S}}\cos(\varphi/2)\) and \(\epsilon_{b\text{F}}=D_{\text{F}}\kappa_{b\text{F}}/d_{\text{F}}\). At \(z=\pm d_{\text{F}}\), the boundary conditions Eq.(\ref{Appendix Strong PE: Boundary Cond.}) are assumed to be identical, but with opposite signs. The Green's functions are diagonal in spin space and the solution to Eq.(\ref{Appendix Strong PE: Integrated Usadel}) is
	\begin{equation}
	\hat{g}_{\text{F},\pm}=g^{(\text{F})}_{\pm}\hat{\tau}_{3}+f^{(\text{F})}_{\pm}\hat{\tau}_{1}
	\label{Appendix Strong PE: Form GF in F}
	\end{equation}
	where the  coefficients are given by \(g^{(\text{F})}_{\pm}=\tilde{\omega}_{\pm}/\tilde{\zeta}_{\tilde{\omega}_{\pm}}(\varphi)\), \(f^{(\text{F})_{\pm}}=\tilde{\Delta}_{\text{F}}(\varphi)/\tilde{\zeta}_{\tilde{\omega}_{\pm}}(\varphi)\) with \(\tilde{\omega}_{\pm }=\tilde{\omega}_{\text{F}}\pm iJ\) and \(\tilde{\zeta}_{\tilde{\omega}_{\pm}}(\varphi)=\sqrt{\tilde{\omega}_{\pm}^{2}+\tilde{\Delta}_{\text{F}}^{2}(\varphi)}\). 
	
	In order to find the induced magnetization in the superconductor S, we suppose that \(\check{g}_{\text{S}}\) deviates weakly from its bulk value Eq.(\ref{Appendix Strong PE: BCS Sol}) and linearize Eq.(\ref{Appendix Strong PE: Usadel S}). Then we can determine a small correction \(\delta\check{g}_{\text{S}}\) to \(\check{g}_{\text{S}}\)
	
	\begin{equation}
	-\partial_{zz}\delta\check{g}_{\text{S}}+\kappa_{\text{S},\omega}^{2}\delta\check{g}_{\text{S}}=2\omega\delta\Delta\qty(\omega\check{X}_{11}-\Delta\check{X}_{30})/D_{\text{S}}\label{Appendix Strong PE: Correction S}
	\end{equation}
	where \(\kappa_{\text{S},\omega}^{2}=2\sqrt{\omega^{2}+\Delta^{2}}/D_{\text{S}}\) and employed the relation
	$\delta\check{g}_{\text{S}}\cdot\check{g}_{\text{S}}^{(0)}+\check{g}_{\text{S}}^{(0)}\cdot\delta\check{g}_{\text{S}}=0$, which follows from the normalization condition.
	The induced magnetization is determined by the component \(\delta g_{33}^{(\text{S})}=\Tr(\check{X}_{33}\delta\check{g}_{\text{S}})/4\) and is given by
	\begin{align}
	M_{\text{S}}(z)&=2\pi i T \mu_{B}\nu_{\text{S}}\sum_{\omega\geq0}\delta g_{33}^{(\text{S})}(\pm d_{\text{F}})\exp(-\kappa_{\text{S},\omega}\abs{z\mp d_{\text{F}}})\notag\\
	&\equiv-\sum_{\omega\geq0}m_{\text{S}}(\varphi)\exp(-\kappa_{\text{S},\omega}\abs{z\mp d_{\text{F}}})
	\label{Eq.1:M(z)}
	\end{align}
\end{widetext}
 In the case of a \textbf{strong PE} \textit{i.e.} (\(R_{b\square}/\varrho_{\text{F}}<\sqrt{D_{\text{F}}/(d_{\text{F}}^{2}J)}\)) (see also Appendix A), it is given by
	\begin{equation}
	m_{\text{S}}(\varphi)=M_{0}\frac{d_{\text{F}}}{\xi_{\text{S}}}\frac{2\pi T}{\Delta}\frac{(\tilde{\omega}^2+1)^{1/4}\cos[2](\varphi/2)}{(\tilde{\omega}^2+\cos[2](\varphi/2))^{3/2}}\label{Eq.2:Strong M(phi)}
	\end{equation}
	where \(D_{\text{F}}\) is the diffusion coefficient of the F film and \(\xi_{\text{S}}^{2}=D_{\text{S}}/2\Delta\), \(\tilde{\omega}=\omega/\Delta\) (see Appendix A). Here, the magnetic moment in the F film \(M_{0}\) is given by \(M_{0}=\mu_{B}\nu_{\text{F}}J\), where \(\mu_{B}\) and \(\nu_{\text{F}}\) are the effective Bohr magneton and the density of states in the F film, respectively.\\
	In the case of a \textbf{weak PE} (\(R_{b\square}/\varrho_{\text{F}}d_{\text{F}}>\sqrt{D_{\text{F}}/(d_{\text{F}}^{2}J)}\)) the calculations are analogous (see Appendix B for details) and one finds
	\begin{align}
	m_{\text{S}}(\varphi)=&4\mu_{B}\nu_{\text{S}}2\pi T G_{\text{S}}F_{\text{S}}^{2}\kappa_{b\text{S}}\kappa_{b\text{F}}\lambda_{\text{S}}\notag\\
	&\times\Im{\frac{\cos[2](\varphi/2)+\sin[2](\varphi/2)\tanh[2](\theta_{\text{F}})}{\kappa_{\text{F},\omega}\tanh(\theta_{\text{F}})}}\label{Eq.3:Weak M(phi)}
	\end{align}
	where \(\theta_{\text{F}}=\kappa_{\text{F},\omega}d_{\text{F}}\), \(\kappa_{\text{F},\omega}^2=(\abs{\omega}+iJ_{\omega})/D_{\text{F}}\), \(J_{\omega}=J\text{sign}(\omega)\) and \(\lambda_{\text{S}}\) is the London penetration depth. Note that in the case of a strong PE the function \(m_{\text{S}}(\varphi)\) does not contain the interface resistance \(R_{b\square}\) while in the case of a weak PE it is proportional to \(R_{b\square}^{-2}\). \\
	In both cases, the induced magnetization in S depends on the phase difference \(\varphi\). The total magnetic moment induced in the two superconductors \(\mathcal{M}_{\text{S}}\) is determined by integration over \(z\) . At zero temperature the summation over the Matsubara frequencies can be replaced by an integral (\(2\pi T \sum_{\omega}(...)\Rightarrow\int_{0}^{\infty}\dd{\omega}(...)\)). In particular we find the following expression for the case of a strong proximity effect
	\begin{equation}
	\mathcal{M}_{\text{S}}=-2d_{\text{F}}M_{0} \qquad \text{for} \quad\varphi\neq\pi
	\end{equation}
    The magnetic moment induced in the superconductors does not depend on the phase difference \(\varphi\) at any \(\varphi\) except the points \(\varphi_{n}=\pi(2n+1)\). It compensates exactly the total magnetic moment of the ferromagnetic film \(\mathcal{M}_{\text{F}}=2d_{\text{F}}M_{0}\). The overall dependence \(\mathcal{M}_{\text{S}}(\varphi)\) as a function of \(\varphi\) is shown in Fig.\ref{fig:2 Phase dependence} for different temperatures \(T\). At temperatures close to the critical temperature \(T_{c}\) the dependence of \(\mathcal{M}_{S}(\varphi)\) is almost sinusoidal. In the case of a weak PE, the total induced magnetization is much less than \(2d_{\text{F}}M_{0}\).
    
	Since the vectors \(\vb{M}_{\text{S}}\) and \(\vb{M}_{\text{F}}\) are aligned in the opposite directions, two identical magnetic granules embedded in a superconductor would interact antiferromagnetically with each other. Indeed, the magnetic moment \(\vb{M}_{\text{S}}=-\vb{M}_{\text{F}1}\), produced by one granule with the magnetic moment \(\vb{M}_{\text{F}1}\) will tend to orient the magnetic moment \(\vb{M}_{\text{F}2}\) of another granule in the direction opposite to \(\vb{M}_{\text{F}1}\). The characteristic length of this interaction is of the order of \(\xi_{\text{S}}\) \cite{bergeret_spin_2004}. There is a similarity between this case and the case considered in Ref. \cite{yao_enhanced_2014}, where it was found that at large distances the interaction between two magnetic impurities in a superconductor is antiferromagnetic one. However, the statement about antiparallel orientation of \(\vb{M}_{\text{S}}\) and \(\vb{M}_{\text{F}}\) is valid only for a diffusive superconductor. In the ballistic case the magnetization \(M_{\text{S}}(z)\) oscillates in space \cite{bergeret_inverse_2005,kharitonov_oscillations_2006}.
	
	The dependence of the induced magnetization on the phase difference, leads to interesting phenomena. For example, if the current \(I\) through the junction exceed the critical value \(I_{c}\), the phase difference increases in time: \(\varphi(t)=2eVt/\hbar\). This means that the induced magnetization \(M_{\text{S}}(t)\) will oscillate in time with the Josephson frequency \(\omega=2eV/\hbar\). Another interesting feature of spin polarization in SF systems is that in the case of a nonuniform magnetization \(\vb{M}_{\text{F}}(\vb{r}_{\perp})\), the magnetization vector \(\vb{M}_{\text{S}}(\vb{r}_{\perp})\) is an inverse mirror image of the vector \(\vb{M}_{\text{F}}(\vb{r}_{\perp})\): \(\vb{M}_{\text{S}}(\vb{r}_{\perp})=-\vb{M}_{\text{F}}(\vb{r}_{\perp})\). This relation is true if the characteristic length of the magnetization variation is much greater than the coherence length in S. For example, in the case of magnetic skyrmions, which can occur as topological structures in the magnetization profile of chiral ferromagnets, a skyrmion with opposite polarity arises in the S film. Since a Josephson junction is one of the most sensitive devices for probing the magnetic field, careful study of the influence of the magnetic field in SFS junctions can lead to the detection of the spin polarization in superconductors S.

	\section{Magnetostatics in SFS junction}
	
	In the following we proceed by considering the magnetostatics of a planar SFS Josephson junction with two identical SF interfaces. The \(z\)-axis is normal to the interface (see Fig.\ref{fig:1 Scheme}) and the thickness of the F film is equal to \(2d_{\text{F}}\). The magnetic properties of the considered system are described be the magnetic induction \(\vb{B}=\curl{\vb{A}}\) and the magnetic field \(\vb{H}\), which are related by the standard relation \(\vb{B}=\vb{H}+4\pi\vb{M}\). The magnetization \(\vb{M}\) exists not only in the F film (\(\vb{M}_{\text{F}}\)), but also in the superconducting region (as the induced magnetization \(\vb{M}_{\text{S}}\)). The magnetic field \(\vb{H}\) obeys 
	\begin{equation}
	\curl{\vb{H}}=\frac{4\pi}{c}\vb{j} \label{Eq.6:Maxwell}
	\end{equation}
	where \(\vb{j}\) is the density of the Meissner current. It is connected to the vector potential \(\vb{A}\) and the phase \(\vb{\chi}\) of the order parameter via the standard gauge invariant expression
	\begin{equation}
	\vb{j}_{\text{S(\text{F})}}=-\frac{c}{4\pi}\qty{\lambda_{\text{S}(\text{F})}^{-2}\qty(\vb{A}-\frac{\Phi_{0}}{2\pi}\grad{\chi})}
	\label{Eq.7:currentphase}
	\end{equation}
	Here, \(\lambda_{\text{S}(\text{F})}\) is the London penetration depths in S and F, respectively, and \(\Phi_{0}=hc/2e\) is the magnetic flux quantum. 
	
	Our goal is to find the relation between an applied magnetic field \(\vb{H}\) and the in-plane gradient \(\grad{\varphi}\) of the phase difference \(\varphi\). 
	We consider an in-plane magnetic field \(\vb{H}=(0,H,0)\) and the in-plane magnetization \(\vb{M}=(0,M(z),0)\) and we further set \(\vb{A}=(A,0,0)\) and \(\vb{j}=(j,0,0)\). Assuming that the \(x\)-dependence of all quantities of interest (\(\vb{H}\) etc.) is weak on the length scale of the London penetration depth \(\lambda_{\text{S}(\text{F})}\) we apply \(\curl\) to the Maxwell equation,  Eq.(\ref{Eq.6:Maxwell}) and obtain the following equation for \(\vb{H}_{\text{S}}\) in the S regions
	\begin{equation}
	\partial_{zz}^{2} H_{\text{S}}(x,z)-\lambda_{\text{S}}^{-2}H_{\text{S}}(x,z)=4\pi\lambda_{\text{S}}^{-2}M_{\text{S}}(z) \label{Eq.8:DGL for Hs}
	\end{equation}
	 with the induced magnetization \(M_{\text{S}}(z)\) given by Eqs.(\ref{Eq.1:M(z)})-(\ref{Eq.2:Strong M(phi)}). Here, we employ the gauge \(\div{\vb{H}}=0\) and use Eq.(\ref{Eq.7:currentphase}) for the current density \(\vb{j}_{\text{S}}\).	Eq.(\ref{Eq.8:DGL for Hs}) is complemented by the boundary condition (see section Appendix C for details)
	\begin{equation}
	\qty[\partial H_{\text{S}}(x)]=\lambda_{\text{S}}^{-2}\qty(\qty[A_{\text{S}}(x)]-\frac{\Phi_{0}}{2\pi}\partial_{x}\varphi(x)) \label{Eq.9: Boundary condition}
	\end{equation}
	where \(\qty[\partial H_{\text{S}}(x)]=\partial_{z}H_{\text{S}}(x,z)|_{z=d_{\text{F}}}-\partial_{z}H_{\text{S}}(x,z)|_{z=-d_{\text{F}}}\), \(\qty[A_{\text{S}}(x)]=A_{\text{S}}(x,d_{\text{F}})-A_{\text{S}}(x,-d_{\text{F}})\) and \(\varphi(x)=\chi(x,d_{\text{F}})-\chi(x,-d_{\text{F}})\) is the phase difference across the junction.
	The difference \(\qty[A_{\text{S}}(x)]\) can be found by taking the continuity of the vector potential at the SF interfaces into account
	\begin{equation}
	\qty[A_{\text{S}}(x)]=\qty[A_{\text{F}}(x)]=2d_{\text{F}}\qty(H_{0}(x)+4\pi M_{\text{F}})\label{Eq.10: Boundary Condition Vectorpotential}
	\end{equation}
	Here, \(H_{0}\) is an integration constant which approximately  coincides with the magnetic field in F. 
	The solution of Eq.(\ref{Eq.8:DGL for Hs}) satisfying boundary condition has the form
	\begin{align}
	H_{\text{S}}(x,z)=&H_{orb}(x)\exp(-\lambda_{\text{S}}^{-1}\abs{z\mp d_{\text{F}}})\notag\\
	&-4\pi\sum_{\omega\geq0}\frac{m_{\text{S}}(\varphi)}{(\lambda_{\text{S}}\kappa_{\text{S},\omega})^2}\exp(-\kappa_{\text{S},\omega}\abs{z\mp d_{\text{F}}})\label{Eq.11: Horb}
	\end{align}
	with 
	\begin{equation}
	H_{orb}(x)=4\pi\sum_{\omega\geq0}\frac{m_{\text{S}}(\varphi)}{(\lambda_{\text{S}}\kappa_{\text{S},\omega})}-\frac{d_{\text{F}}}{\lambda_{\text{S}}}\qty(H_{0}(x)+4\pi M_{0})+\frac{\Phi_{0}}{4\pi\lambda_{\text{S}}}\partial_{x}\varphi
	\end{equation}
	The short-ranged component, which decreases over the superconducting coherence length \(\xi_{\text{S}}\), is a direct consequence of the inverse proximity effect. In addition, the spin polarization in the S results in a modification of the long-ranged component of the magnetic field (see Eq.(\ref{Eq.11: Horb})) which decays over the London penetration depth \(\lambda_{\text{S}}\). Its amplitude \(H_{orb}\) determines the magnetic field caused by orbital motion of the condensate (Meissner currents). The quantity \(\kappa_{\text{S},\omega}\) is defined in Eq.(\ref{Appendix Strong PE: Correction S}).
	Note that the field \(H_{0}(x)\) and the phase difference \(\varphi\) are smoothly varying function of \(x\). In the following, we assume that \(\xi_{\text{S}}\ll\lambda_{\text{S}}\) \textit{i.e.} neglecting the terms of the order of \(\order{\xi_{\text{S}}/\lambda_{\text{S}}}\), so that \(H_{\text{S}}(x,z)\) is dominated by the orbital contribution.\\
	Next, we use the continuity condition for the field \(H_{\text{S(\text{F})}}\) at at the interfaces \mbox{\(z=\pm d_{\text{F}}\)} \textit{i.e.} \mbox{\(H_{\text{S}}(x,\pm d_{\text{F}})=H_{\text{F}}(x,\pm d_{\text{F}})\)} with \mbox{\(H_{\text{F}}(x,z)=H_{0}(x)+(H_{0}(x)+4\pi M_{0})\lambda_{\text{F}}^{-2}z^2/2\)}, and arrive to the following equation for \(\varphi\)
	\begin{equation}
	\partial_{\tilde{x}}\varphi(\tilde{x})=2\pi\qty[\tilde{\Phi}_{m}(\tilde{x})-p_{\varphi}]\label{Eq.13: DGL phase}
	\end{equation}
	where \(\tilde{x}=x/L\).\\
	 \(\tilde{\Phi}_{m}(\tilde{x})\) and \(p_{\varphi}\) are defined as
	\begin{align}
	\tilde{\Phi}_{m}(\tilde{x})=& \qty{H_{0}(\tilde{x})L\qty(2\lambda_{\text{S}}+2d_{\text{F}})+4\pi M_{0}L2d_{\text{F}}}/\Phi_{0}\\
	p_{\varphi}=&2\gamma_{\varphi}\qty{4\pi M_{0}L2d_{\text{F}}}/\Phi_{0} \label{Eq.15: p}
	\end{align}
	In simple words, \(\tilde{\Phi}_{m}\) and \(p_{\varphi}\) are the normalized magnetic flux in the junction and a normalized \(\varphi\)-dependent contribution caused by spin polarization in S, respectively. The coefficient \(\gamma_{\varphi}\) is given by
	\begin{equation}
	\gamma_{\varphi}=\frac{\xi_{\text{S}}}{2d_{\text{F}}M_{0}}\sum_{\omega\geq0}\frac{m_{\text{S}}(\varphi)}{(\tilde{\omega}^{2}+1)^{1/4}}\label{Eq.16:Gamma}
	\end{equation}
	with \(\xi_{\text{S}}=\sqrt{D_{\text{S}}/2\Delta}\). The function \(\gamma_{\varphi}\) is a periodic function of \(\varphi\) and has different form in the limit of a strong and weak PE, see Appendices A and B for details. For temperatures, \(T\), close to \(T_{c}\), we get \(\gamma_{\varphi}\cong\gamma_{0}\cos[2](\varphi/2)\), with \(\gamma_{0}\) being a constant. If one considers \(\gamma_{0,\varphi}=0\) \textit{i.e.} no spin polarization, Eq.(\ref{Eq.13: DGL phase}) coincides with the well known equation derived by Ferrel and Prange \cite{ferrell_self-field_1963} (see also \cite{kulik_josephson_1972,likharev_superconducting_1979,barone_physics_1982}). Then, the right-hand side of Eq.\ref{Eq.13: DGL phase} is the normalized magnetic flux in the junction \(\tilde{\Phi}_{m}\) \textit{i.e.} the flux related to the magnetic inductance \(B(x)\). It consists of an external field \(H_{ext}\),  the magnetic field created by the Josephson current \(I_{J}\) and the total magnetic moment of the F and the S. In the case of low barrier resistance and low temperatures \(\gamma_{0}=1\), \textit{i.e.} the flux coincides with the magnetic flux in the junction caused by an external magnetic field \(H_{ext}\) as the magnetic flux in the F given by \(4\pi M_{0} 2d_{\text{F}}L\) is compensated by the magnetic flux in the S.
		
	Another useful relation between \(\tilde{\Phi}_{m}\) and \(\varphi\) can be obtained from the Maxwell equation
	\begin{equation}
	\pdv{H_{0}(x)}{
		x}=\frac{4\pi}{c}j_{c}\sin(\varphi)
	\end{equation}
	which after some straightforward algebra can be written as
	\begin{equation}
	\pdv{\tilde{\Phi}_{m}(\tilde{x})}{
		\tilde{x}}=r\sin(\varphi)  \label{Eq.18: DGL FLUX}
	\end{equation}
	with \(r=(4\pi L^{2} j_{c}/c)(2\lambda_{\text{S}}+2d_{\text{F}})/\Phi_{0}\). One can easily find the critical current \(j_{c}\) in the case of a strong and a weak PE, which is given in the Appendix C in terms of microscopic parameters.\\
	Finally, Eqs.(\ref{Eq.13: DGL phase})-(\ref{Eq.18: DGL FLUX}) determine the relation between the "magnetic" flux \(\tilde{\Phi}_{m}\) and the phase difference \(\varphi(x)\) in the presence of an induced magnetization \(M_{\text{S}}(x)\). In principle, the non-linear Eqs.(\ref{Eq.13: DGL phase},\ref{Eq.18: DGL FLUX}) can only be solved numerically. But we can still obtain analytic formulas assuming that the temperatures \(T\) are close to \(T_{c}\). The obtained results remain qualitatively unchanged for any \(T\leq T_{c}\). Near \(T_{c}\)  Eq.(\ref{Eq.13: DGL phase}) acquires the form
	\begin{equation}
	\pdv{\varphi(\tilde{x})}{\tilde{x}}=2\pi\qty[\tilde{\Phi}_{m}(\tilde{x})-p\cos[2](\varphi/2)]\label{Eq.19: DGL PHI}
	\end{equation}
	Here, the coefficient \(p=2\gamma_{0}\qty{4\pi M_{0}L2d_{\text{F}}}/\Phi_{0}\) is directly reflecting the strength of induced magnetic polarization effect in the S and can be easily expressed using Eq.(\ref{Eq.15: p}) and Eq. (\ref{Eq.16:Gamma}) with the function \(\gamma_{0}\) given in Eqs.(\ref{Appendix Strong PE: Gamma})-(\ref{Appendix Weak PE: Gamma}) via \(\gamma_{0}=\gamma_{\varphi}/\cos[2](\varphi/2)\)(see Appendix C). The coefficient \(r\), defined above, depends on the critical current \(j_{c}\) and its form is given explicitly in the Appendix C as well. As we noted above, Eqs.(\ref{Eq.18: DGL FLUX}),(\ref{Eq.19: DGL PHI}) can be solved numerically yet  analytical expression in the limiting cases of $p<<r$  and $p>>r$ can also be obtained. Most importantly,  to estimate how small \(p\) is with respect to \(r\), we rescale the quantities \(p\), \(p_{\varphi}\), \(\tilde{\Phi}_{m}\) and $x$: \(p_{r}=p/\sqrt{r}\), \(\tilde{\Phi}_{m,r}(x)=\tilde{\Phi}_{m}/\sqrt{r}\) and \(X=(x/L)\sqrt{r}\) yielding
	\begin{align}
	\tilde{\Phi}_{m,r}(X)=&\partial_{X}\varphi(X)+p_{r}\cos[2](\varphi(X)/2)\\
	\partial_{X}\tilde{\Phi}_{m,r}(X)=&\sin(\varphi(X))
	\end{align}
	One can see that if \(p_{r}\ll 1\), that is, \(p\ll \sqrt{r}\), the spin polarization in the S can be considered as a small correction. In the opposite limit, \(p\gg\sqrt{r}\), the spin polarization of Cooper pairs leads to drastic changes, for example in the Fraunhofer pattern as we show below.
	
	In the following we estimate parameters of the Josephson Nb/\(\text{Cu}_{0.47}\text{Ni}_{0.53}\)/Nb junctions studied experimentally \cite{kontos_josephson_2002}. These parameters can be readily evaluated for this system as \(2d_{\text{F}}=15 \mbox{-} 25\text{nm}\), \(j_{c}=2\cdot10^{4}\mbox{-}2\cdot10^{3}\text{A}/\text{cm}^{2}\), the total interface resistance \(R_{b}=30\mu\Omega\) for dimensions \(L_{x}\times L_{y}=10\times 10\mu\text{m}^{2}\),  \(R_{b\square}=3\cdot 10^{-11}\Omega\cdot\text{cm}^{2}\), \(D_{\text{F}}=5\text{cm}^{2}/s\), \(\varrho_{\text{F}}=60\mu\Omega\cdot\text{cm}\), , \(J=E_{ex}=850\text{K}\). Thus one finds \(\varrho_{\text{F}}\cdot d_{\text{F}}=6\cdot10^{-11}\Omega\cdot\text{cm}^{2}\geq R_{b\square}\) and the coherence length is \(\xi_{\text{F}}=2.16\text{nm}\). For these parameters the factor \(p\) appears to be small compared to \(r\), so that the induced magnetization \(M_{\text{S}}\) leads to rather small changes in the Josephson effect. For example, \(p\sim30\sqrt{I_{\theta}}\approx 0.3\) for \(I_{\theta}\approx10^{-4}\) and \(p\sim30\sqrt{I_{\theta}}\approx3\) for \(I_{\theta}\approx10^{-2}\); here \(I_{\theta}=\exp(-2\theta_{\text{F}})/\theta_{\text{F}}\). 
	Nevertheless, it is instructive to investigate the Fraunhofer pattern for both small and large values of \(p/r\) having in mind that in some other experimental realization of the SFS junction, the coeficient $p$ can be potentially larger. 
	\begin{figure*}[tbp]
		\includegraphics[width=\linewidth]{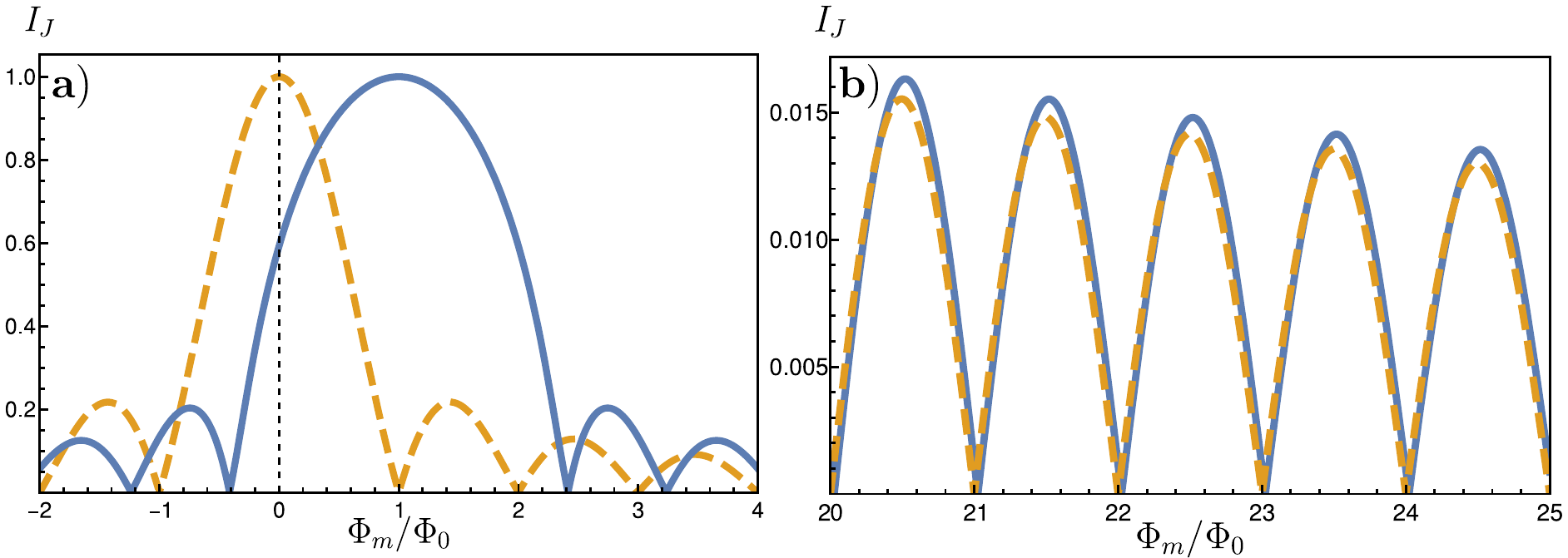}
		\caption{(Color online) a) The dependence of the critical current \(\abs{\tilde{I}_{J}(\tilde{\Phi}_{m})}\) for \(p=2\) (solid blue curve). The dashed (orange) line refers to the Fraunhofer pattern \(\abs{\tilde{I}_{Fr}(\tilde{\Phi}_{m})}\) in the absence of of the spin polarization (SP) \label{fig:3 New Fraunhofer} b) shows the same curves as in (a) but for larger values of \(\Phi_{m}/\Phi_{0}\). One can see that the deviations from the standard Fraunhofer pattern becomes smaller with increasing \(\Phi_{m}/\Phi_{0}\).\label{fig:5 New Fraunhofer large}}
	\end{figure*}

	\section{Fraunhofer pattern}
	
	In this section, we begin with a brief discussion of  the case where  spin polarization in S leads to small corrections to the standard Fraunhofer pattern. Most of the remaining section is devoted towards the opposite case where the spin polarization plays a dominate role, resulting in a drastic change in the Fraunhofer pattern.
	
	\subsection*{Weak spin polarization: \(\mathbf{p\ll\sqrt{r}}\)}
	Here, the phase difference \(\varphi(\tilde{x})\) can be represented by: \(\varphi(\tilde{x})=\varphi_{0}(\tilde{x})+\varphi_{1}(\tilde{x})+\varphi_{2}(\tilde{x})\), where \(\varphi_{0}(\tilde{x})=\tilde{\Phi}_{m}\tilde{x}+c\) is an arbitrary constant determined by the requirement to maximize the current \(I_{J}\). The normalized flux is given by \(\tilde{\Phi}_{m}=\tilde{\Phi}_{m,0}+\tilde{\Phi}_{m,1}+\tilde{\Phi}_{m,2}\). In zero-order approximation the magnetic flux created by the current \(I_{J}\) can be neglected and therefore \(\tilde{\Phi}(\tilde{x})=\tilde{\Phi}_{ext}+\tilde{\Phi}_{\text{F}}+\tilde{\Phi}_{\text{S}}\). The constant \(\tilde{\Phi}_{ext}\), \(\tilde{\Phi}_{\text{F}}\) and \(\tilde{\Phi}_{\text{S}}\) are defined as: \(\tilde{\Phi}_{ext}=2\pi H_{ext}L(2d_{\text{F}}+2\lambda_{\text{S}})\), \(\tilde{\Phi}_{\text{F}}=2\pi(4\pi M_{0}L 2d_{\text{F}})/\Phi_{0}\) and \(\tilde{\Phi}_{\text{S}}=2\pi(4\pi M_{0}L 2d_{\text{F}}2\gamma_{0})/\Phi_{0}\). Expanding \(\varphi(\tilde{x})\) and \(\tilde{\Phi}(\tilde{x})\) we find the current
	\begin{equation}
	\tilde{I}_{\text{max}}=\frac{\sin(\pi\tilde{\Phi}_{m})}{\pi\tilde{\Phi}_{m}}+\delta\tilde{I}
	\end{equation}
	Here the correction \(\delta\tilde{I}\) contains small terms of the order \(\tilde{p}^{2}/(2\Phi_{m,0}^{3})\) as well as \(\sin(2\pi\tilde{\Phi}_{m})\), \(\sin(3\pi\tilde{\Phi}_{m})\) etc., where \(\tilde{p}\equiv(p+r/\tilde{\Phi}_{m,0})\) (see Appendix D for further details). At large \(H_{ext}\) \textit{i.e.} at large \(\tilde{\Phi}_{m}\) the contribution from the spin polarization \(p\) will dominated in \(\tilde{p}\).
	
	\subsection*{Strong spin polarization: \(\mathbf{p\gg\sqrt{r}}\)}
	
	In this case, our main approximation is a vanishing \(r\), such that Eq.\ref{Eq.19: DGL PHI} is described by a constant effective flux \(\tilde{\Phi}_{m}(x)=\tilde{\Phi}_{m}\). Now, we can write the normalized current \(\tilde{I}_{J}=I/Lj_{c}\) as
	\begin{equation}
		\tilde{I}_{J}=\frac{1}{L}\int_{0}^{L}\dd{x}\sin(\varphi(x))=\int_{\varphi_{0}}^{\varphi_{L}}\frac{\dd{\varphi}}{2\pi}\frac{\sin(\varphi)}{\tilde{\Phi}_{m}-p\cos[2](\varphi/2)}\label{Eq.23 :Current small}
	\end{equation}
	with \(\varphi_{L}=\varphi(L)\) and \(\varphi_{0}=\varphi(0)\).\\
	\begin{figure}[tbp]
	\includegraphics[width=\columnwidth]{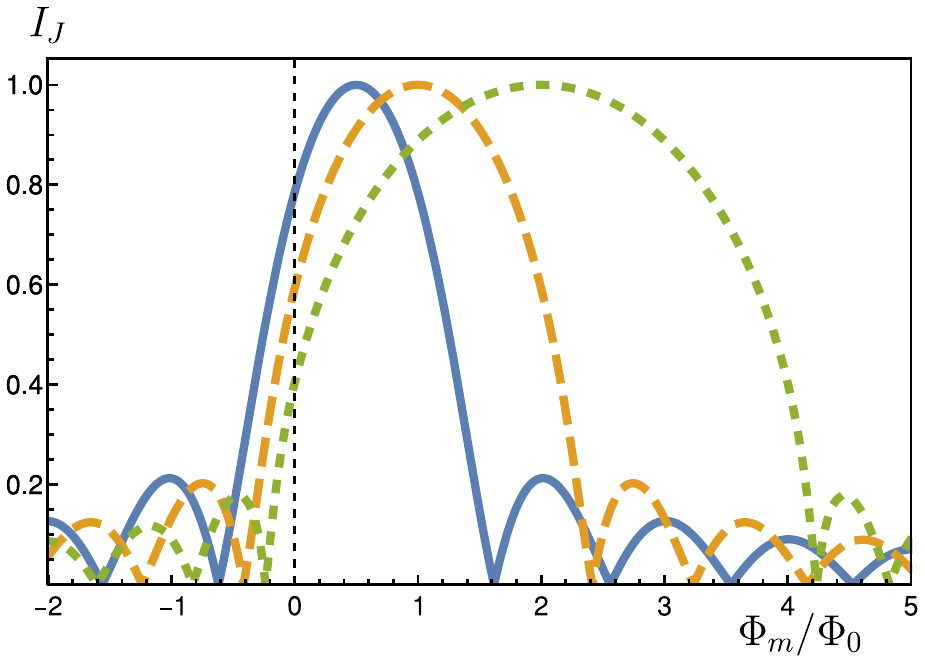}
	\caption{(Color online) Comparison of the dependence of the critical current \(\abs{\tilde{I}_{J}(\tilde{\Phi}_{m})}\) for \(p=1\) (blue solid line), \(p=2\) (orange dashed line) and \(p=4\) (green dotted line) .\label{fig:4 New Fraunhofer diff p}}
\end{figure}
	After integration, we obtain
	\begin{align}
		\tilde{I}_{J}=&\frac{1}{\pi p}\ln(\frac{p-2\tilde{\Phi}_{m}+p\cos(\varphi_{L})}{p-2\tilde{\Phi}_{m}+p\cos(\varphi_{0})})\notag\\
		=&\frac{1}{\pi p}\ln(\frac{a^2+T_{L}^2}{a^2+T_{0}}\frac{1+T_{0}^2}{1+T_{L}^2}) \label{Eq.24: Current with TL}
	\end{align}
	where \(T_{L}=\tan(\varphi_{L}/2)\), \(T_{0}=\tan(\varphi_{0}/2)\) and \(a^2=(\tilde{\Phi}_{m}-p)/\tilde{\Phi}_{m}\). The quantities \(T_{L}\) and \(T_{0}\) are coupled via
	\begin{equation}
		L=\int_{0}^{L}\dd{x}=L\int_{\varphi_{0}}^{\varphi_{L}}\frac{\dd{\varphi}}{2\pi}\frac{1}{\tilde{\Phi}_{m}-p\cos[2](\varphi/2)}
	\end{equation}
	Calculating the integral, we obtain the relation
	\begin{equation}
		T_{L}=\frac{T_{0}+ba^2}{1+bT_{0}} \label{Eq.26: TL}
	\end{equation}
	with \(b=a^{-1}\tan(\pi a \tilde{\Phi}_{m})\).	So far, Eq.(\ref{Eq.24: Current with TL},\ref{Eq.26: TL}), determine the critical current \(\tilde{I}_{J}\) as a function of an arbitrary constant \(T_{0}\equiv\tan(\varphi_{0}/2)\). In order to find the maximum of the current \(\tilde{I}_{J}\), we need to determine the value of \(T_{0}\) which maximizes the current. The roots of the equation \(\partial \tilde{I}_{J}/\partial T_{0}=0\) are given by
	\begin{align}
		T_{m}=&\frac{1}{2a}\left(-(1+a^2)\tan(\pi a\tilde{\Phi}_{m})\notag\right.\\
		&\left.\pm\sqrt{4a^2+(1+a^2)^2\tan[2](\pi a\tilde{\Phi}_{m})}\right)\label{Eq.27: Roots}
	\end{align}
	Substituting \(T_{m}\) into Eq.(\ref{Eq.24: Current with TL}) and using Eq.(\ref{Eq.26: TL}), we can extract the solution for the maximal current \(\tilde{I}_{J}(\tilde{\Phi}_{m})\). The result is shown in Fig.\ref{fig:3 New Fraunhofer} for \(p=2\) characterizing a finite contribution from the spin polarization. The dependence \(\tilde{I}_{J}(\tilde{\Phi}_{m})\) is compared with the standard Fraunhofer pattern in the absence of the spin polarization 
	\begin{equation}
		I_{Fr}(\Phi_{m}/\Phi_{0})=\abs{\frac{\sin(\pi\Phi_{m}/\Phi_{0})}{\pi\Phi_{m}/\Phi_{0}}}
		\label{Eq.28: Fraunhofer without SP}
	\end{equation}
	 where \(\Phi_{m}=L\qty{H_{0}\qty(2\lambda_{\text{S}}+2d_{\text{F}})+4\pi M_{0}2\lambda_{\text{S}}}\).
	 \begin{figure}[tbp]
	 	\includegraphics[width=\columnwidth]{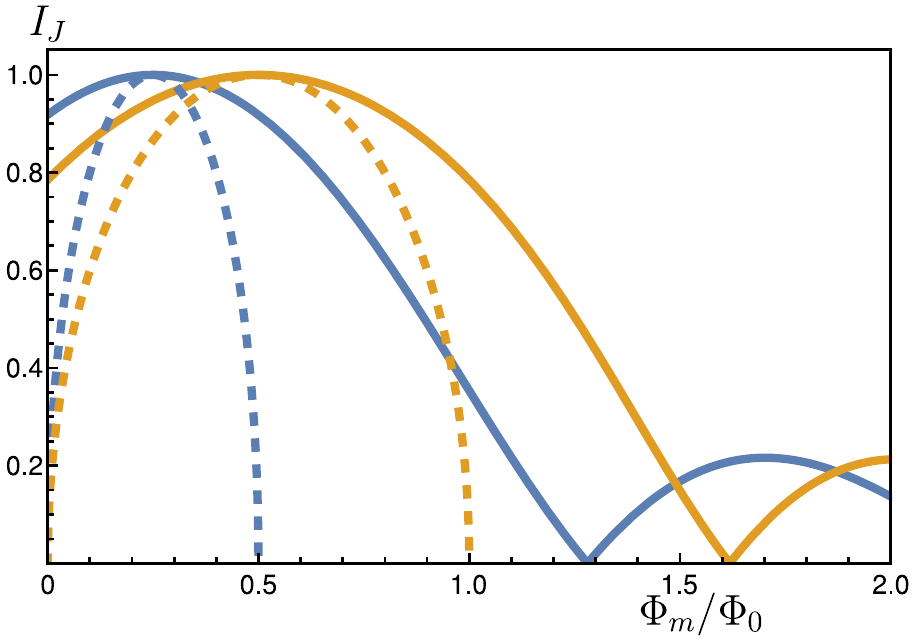}
	 \caption{(Color online) Comparison of the solution of the critical current \(\abs{\tilde{I}_{J}(\tilde{\Phi}_{m})}\) in Eq.(\ref{Eq.24: Current with TL}) (solid lines) and \(\tilde{I}_{J,ad}=\sqrt{1-(2\tilde{\Phi}_{m}-p)^2/p^{2}}\) (dashed lines) for \(p=0.5\) (blue line) and \(p=1\) (orange line).\label{fig:6 Comparison solution}}
	 \end{figure}
	 One can see that for small values of \(p\), the behavior of the critical current \(\abs{I_{J}(\tilde{\Phi}_{m})}\) resembles the shape of the classical Fraunhofer pattern, whereas for increasing \(p\), the difference become more pronounced (see also Fig.\ref{fig:4 New Fraunhofer diff p}). In particular, the induced spin polarization causes a shift of the global maximum of the critical current by an amount of \(p/2\). This shift occurs in addition to the displacement of the global maximum caused by the magnetization in the ferromagnet(see Refs. \cite{banerjee_reversible_2014,glick_critical_2017,satchell_supercurrent_2018}), so that its position is effectively changed by the amount \(4\pi M_{0}L2d_{\text{F}}\qty[1-\gamma_{0}]\). Notice that for \(\gamma_{0}=1\), both displacements cancel each other, so there is no shift.\\
	 Most importantly, the spin polarization causes a broadening of the peaks of the Fraunhofer pattern. This effect is most pronounced for the main maximum (see Fig.\ref{fig:3 New Fraunhofer}). In contrast to the shift, the broadening is only determined by the strength of the induced spin polarization and is as such a direct consequence of the inverse proximity effect. The broadening is stronger for larger \(p\) (see Fig.\ref{fig:6 Comparison solution})\\
	 If one compares the changes of the Fraunhofer pattern for large values \(\Phi_{m}/\Phi_{0}\) (see Fig.\ref{fig:5 New Fraunhofer large}b), one recognizes that the differences to the standard pattern disappear.
	 
	 It should be noted that for \(\tilde{\Phi}_{m}\leq p\), one can obtain another solution for Eq.(\ref{Eq.19: DGL PHI}) with a space-independent phase difference \(\varphi_{ad}\)
	 \begin{equation}
	 	\cos(\varphi_{ad})=(2\tilde{\Phi}_{m}-p)/p
	 \end{equation}
	However, the current \(\tilde{I}_{J,ad}=\sqrt{1-(2\tilde{\Phi}_{m}-p)^2/p^{2}}\) corresponding to this solution is less than the current \(\tilde{I}_{J}\) given by Eq.(\ref{Eq.24: Current with TL}) (see Fig.\ref{fig:6 Comparison solution}). Thus, the Josephson energy \(E_{J,ad}\sim I_{J,ad}(1-\cos(\varphi_{ad}))\) is somewhat higher than the Josephson energy related to the current \(\tilde{I}_{J}\). Since the difference between these currents is small, transitions between two different solution for the current \(\tilde{I}_{J}\) are possible at small \(\tilde{\Phi}_{m}\). Note that some instabilities in the dependence of \(I_{J}(H_{ext})\) are observed in the experiment \cite{banerjee_reversible_2014,glick_critical_2017,satchell_supercurrent_2018}.\\
	One can easily show that for \(p\rightarrow0\), the dependence \(\abs{\tilde{I}_{J}(\tilde{\Phi}_{m})}\) is reduced to \(\abs{\tilde{I}_{Fr}(\tilde{\Phi}_{m})}\). In this limit we have \(a\cong1-2p/\tilde{\Phi}\) and Eq.(\ref{Eq.24: Current with TL}) takes the form
	\begin{equation}
		\tilde{I}_{J}(\tilde{\Phi}_{m})\approx\frac{1}{\pi\tilde{\Phi}_{m}}\qty[\frac{1}{1+T_{0}^{2}}-\frac{1}{1+T_{L}^{2}}]
	\end{equation}
	Substituting Eq.(\ref{Eq.26: TL},\ref{Eq.27: Roots}) with \(a=1\), we obtain Eq.(\ref{Eq.28: Fraunhofer without SP}).
	 \begin{figure}[tbp]
	 	\includegraphics[width=\columnwidth]{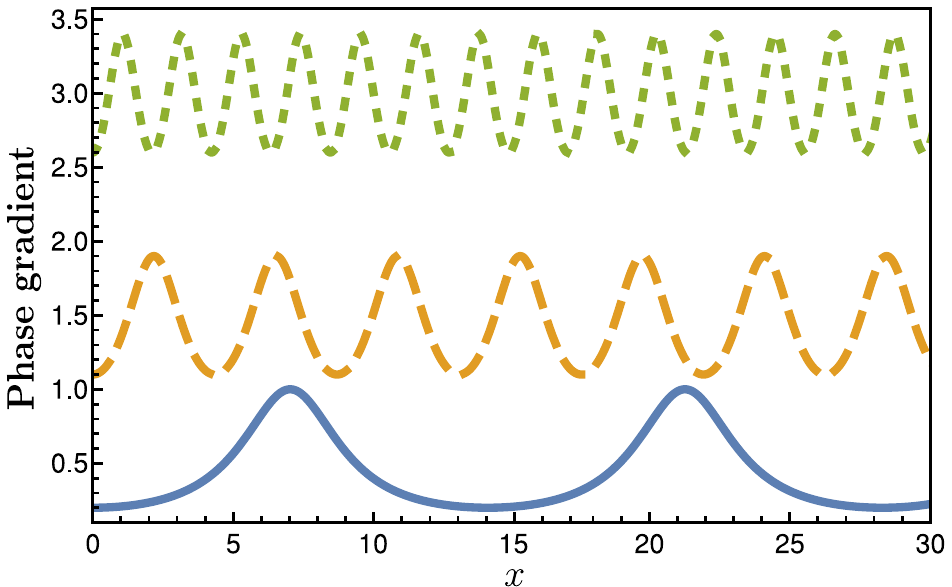}
	\caption{(Color online) Coordinate dependence of the phase gradient \(\partial\varphi(\tilde{x})/\partial\tilde{x}\) for different \(\tilde{\Phi}_{m}\) but fixed \(p=0.4\): \(\tilde{\Phi}_{m}=0.6\) (blue line),  \(\tilde{\Phi}_{m}=1.5\) (orange line) and  \(\tilde{\Phi}_{m}=3\) (green line).\label{fig:7 Phasegradient}}
	 \end{figure}
	We also provide an expression for the coordinate dependence for the phase difference \(\varphi\) which can be obtained from Eq.(\ref{Eq.19: DGL PHI})
	\begin{equation}
		\pdv{\varphi}{\tilde{x}}=\frac{\tilde{\Phi}_{m}^{2}-p^{2}}{\tilde{\Phi}_{m}+p\cos(\kappa(\tilde{x}-\tilde{x}_{0}))}
	\end{equation}
	where \(\kappa=\sqrt{\tilde{\Phi}_{m}^{2}-p^{2}}\). This dependence describes a fluxon in a long  SFS Josephson junction \textit{i.e.} \(L\gg\lambda_{J}\) with the Josephson length \(\lambda_{J}\) (see Fig.\ref{fig:7 Phasegradient}). One can see that at \(\tilde{\Phi}_{m}\) close to \(p\), the fluxons have the form of narrow spikes. At \(\tilde{\Phi}_{m}\gg p\), this dependence has the form of a sinusoidal function. Note that the dependence \(\partial\varphi(\tilde{x})/\partial\tilde{x}\) given by this equation coincides with the temporal dependence of the voltage \(V(t)\) at different currents \(I\) in a point superconducting contact if the following replacements are made: \(V\rightarrow\partial\varphi(\tilde{x})/\partial\tilde{x}\); \(I/I_{c}\rightarrow\tilde{\Phi}_{m}\); \(t\rightarrow\tilde{x}\); \(p=1\) (\(I_{c}\) is the critical current)\cite{Larkin1969}.
	
	\section{Conclusion}
	
	In this manuscript, we studied the influence of the spin polarization of Cooper pairs in superconductors on the Josephson effect in SFS junctions. The expression for the induced magnetization \(M_{\text{S}}\) was obtained in the cases of low and large SF interface resistances \(R_{b\square}\). We have shown that the magnetization \(M_{\text{S}}\) depends on the phase difference \(\varphi\) so that for \(I>I_{c}\) it oscillates in time with the Josephson frequency \(\omega=2eV/\hbar\). The induced spin polarization in the S, contributes to the orbital motion of the condensate (Meissner currents), resulting in a long-ranged magnetic field contribution penetrating the superconductor over the length scale of the London penetration depth \(\lambda_{\text{S}}\). Since the magnetic flux in the junction is not only determined by an applied external magnetic field \(\vb{H}_{ext}\) but also by the total magnetic moment \(\vb{M}_{tot}=\vb{M}_{\text{F}}+\vb{M}_{\text{S}}\), the Fraunhofer pattern depends on the induced magnetization \(\vb{M}_{\text{S}}\). In particular, at low \(R_{b\square}\) and low temperatures \(T\), the total magnetization may turn to zero: \(\vb{M}_{\text{F}}+\vb{M}_{\text{S}}=0\) (full magnetic screening).  With a suitable choice of parameters the Fraunhofer pattern modifies drastically. For example, the spin polarization in the S causes a shift of the Fraunhofer pattern, which is opposite to the displacement by the magnetization in the F. However, even more significant is a broadening of the Fraunhofer peaks, which is particularly pronounced for the peak corresponding to the global maximum. These changes are most notable for a small magnetic flux, such that one should look for features of the spin polarization in the first series of peaks.  Thus we conclude that a careful analysis of the Josephson effect in SFS structures (for example, of the dependence of \(I_{J}(H_{ext})\)) may reveal the presence of the spin polarization in superconductors.
	
	\section{Acknowledgement} The authors acknowledge support from the Deutsche Forschungsgemeinschaft Priority Program SPP2137,
	Skyrmionics, under Grant No. ER 463/10.
	
	\bibliography{literature} 
	
	\onecolumngrid
	\appendix
	\setcounter{equation}{0}
	\setcounter{figure}{0}
	\setcounter{table}{0}
	\makeatletter
	\renewcommand{\theequation}{A\arabic{equation}}
	\renewcommand{\thefigure}{A\arabic{figure}}
	\renewcommand{\bibnumfmt}[1]{[A#1]}


\section{Strong Proximity Effect}

In the following we present several limiting cases for the magnetization in the superconductor for a strong proximity effect. We start by multiplying Eq.(\ref{Appendix Strong PE: Correction S}) by \(\check{X}_{33}\) and calculate the trace. Then, the right-hand side of Eq.(\ref{Appendix Strong PE: Correction S}) is zero and we find a solution
\begin{equation}
\delta g_{33}^{(\text{S})}(z)=\delta g_{33}^{(\text{S})}(\pm d_{\text{F}})\exp(-\kappa_{\text{S},\omega}\abs{z\mp d_{\text{F}}})
\end{equation}
The integration constant is found from the boundary condition Eq.(\ref{Appendix Strong PE: Boundary Cond.}) that yields
\begin{equation}
\partial_{z} \delta g_{33}^{(\text{S})}(z)|_{z=\pm d_{\text{F}}}=\mp 2\kappa_{b\text{S}}F_{\text{S}}\qty(F_{\text{S}}g_{33}^{(\text{F})}-G_{\text{S}}g_{13}^{(\text{F})}\cos(\varphi/2))
\end{equation}
where \(g_{13}^{(\text{F})}=\epsilon_{b\text{F}}i\Im{1/\tilde{\zeta}_{\tilde{\omega}_{+}}(\varphi)}\) and \(g_{33}^{(\text{F})}=i\Im{\tilde{\omega}_{+}/\tilde{\zeta}_{\tilde{\omega}_{+}}(\varphi)}\). The function \(\tilde{\zeta}_{\tilde{\omega}_{\pm}}(\varphi)\) is: \(\tilde{\zeta}_{\tilde{\omega}_{\pm}}(\varphi)=\sqrt{\tilde{\omega}_{\pm}^{2}+\tilde{\Delta}_{\text{F}}^{2}(\varphi)}\).\\
We obtain for \(\delta g_{33}^{(S)}(z)\)
\begin{equation}
\delta g_{33}^{(S)}(z)=\frac{2\kappa_{b\text{S}}}{\kappa_{\text{S},\omega}}F_{\text{S}}\qty(F_{\text{S}}g_{33}^{(F)}-G_{\text{S}}g_{13}^{(F)}\cos(\varphi/2))\exp(-\kappa_{\text{S},\omega}\abs{z\mp d_{\text{F}}})=\delta g_{33}^{(S)}(\pm d_{\text{F}})\exp(-\kappa_{\text{S},\omega}\abs{z\mp d_{\text{F}}})\label{Appendix Strong PE: Solution 33 component S}
\end{equation}
One can see that this correction \(\delta\check{g}_{\text{S}}\) is small if the condition \(\kappa_{b\text{S}}\xi_{\text{S}}\ll1\) is fulfilled, that is, \(R_{b\square}\gg\varrho_{\text{S}}\xi_{\text{S}}\).\\
Consider the case where \(\Delta\ll J\ll\epsilon_{b\text{F}}\). In this limit \(\tilde{\omega}_{+}\approx\epsilon_{b\text{F}}(G_{\text{S}}+iJ)\) and \(\tilde{\zeta}_{\tilde{\omega}_{+}}(\varphi)=\epsilon_{b\text{F}}\zeta_{\varphi}(1+i\tilde{J}G_{\text{S}}/\zeta_{\varphi}^{2})\). The function \(g_{13}^{(\text{F})}\) and \(g_{33}^{(\text{F})}\) are equal to
\begin{align}
g_{13}^{(\text{F})}=&-\frac{i\tilde{J}}{\zeta_{\varphi}}\frac{G_{\text{S}}F_{\text{S}}\cos(\varphi/2)}{\zeta_{\varphi}^{2}}\\
g_{13}^{(\text{F})}=&\frac{i\tilde{J}}{\zeta_{\varphi}}\frac{F_{\text{S}}^{2}\cos[2](\varphi/2)}{\zeta_{\varphi}^{2}}
\end{align}
with \(\tilde{J}=J/\epsilon_{b\text{F}}\), \(\zeta_{\varphi}=\sqrt{G_{\text{S}}^{2}+F_{\text{S}}\cos[2](\varphi/2)}\). By using Eq.(\ref{Appendix Strong PE: Solution 33 component S}), we obtain for \(\delta g_{33}^{(\text{S})}(\pm d_{\text{F}})\)
\begin{equation}
\delta g_{33}^{(\text{S})}(\pm d_{\text{F}})=\frac{i\tilde{J}\kappa_{b\text{S}}}{\kappa_{\text{S},\omega}}\frac{2F_{\text{S}}^{2}\cos[2](\varphi/2)}{\zeta_{\varphi}}
\end{equation}

The magnetization \(M_{\text{S}}(z)\) in Eq.(\ref{Eq.1:M(z)}) is expressed through the function \(m_{\text{S}}(\varphi)\) that is given in Eq.(\ref{Eq.2:Strong M(phi)}). Near \(T_{c}\) we obtain for \(\gamma_{\varphi}\) (see Eq.(\ref{Eq.16:Gamma}))

\begin{equation}
\gamma_{\varphi}=\qty(\frac{\Delta}{\pi T})^{2}2\cos[2](\varphi/2)\sum_{n\geq 0} (2n+1)^{-3} \label{Appendix Strong PE: Gamma}
\end{equation}

\setcounter{equation}{0}
\setcounter{figure}{0}
\setcounter{table}{0}
\makeatletter
\renewcommand{\theequation}{B\arabic{equation}}
\renewcommand{\thefigure}{B\arabic{figure}}
\renewcommand{\bibnumfmt}[1]{[B#1]}

	\section{Weak Proximity Effect}

	Consider an SFS junction with a high interface resistance so that only a weak PE occurs. Then the condensate wave function \(\check{f}\) is small and we can linearized Eq.(\ref{Appendix Strong PE: Usadel S}). In the F film, the function \(\check{f}_{\text{F}}\)\ obeys the equation
	\begin{equation}
	\partial_{zz}^{2}\hat{f}_{\text{F}\pm}-\kappa_{\text{F}\pm}^{2}\hat{f}_{\text{F}\pm}=0, \quad \text{F film}  \label{Appendix Weak PE: Linearized}
	\end{equation}
	where \(\kappa_{\text{F}\pm}^{2}=2(\abs{\omega}\pm i J_{\omega})/D_{\text{F}}\) with \(J_{\omega}=J\text{sign}(\omega)\). In zeroth order approximation, the Green's functions in the S films have the form
	\begin{equation}
	\hat{g}_{\text{S}}^{(0)}(\pm d_{\text{F}})=G_{\text{S}}\hat{\tau}_{3}+F_{\text{S}}\qty{\cos(\varphi/2)\hat{\tau}_{1}\pm\sin(\varphi/2)\hat{\tau}_{2}}
	\end{equation}	
	The solution of Eq.(\ref{Appendix Weak PE: Linearized}) is
	\begin{equation}
	\hat{f}_{\text{F}\pm}(z)=\hat{C}_{\pm}\cosh(\kappa_{\text{F}\pm}z)+\hat{S}_{\pm}\sinh(\kappa_{\text{F}\pm}z)\label{Appendix Weak PE: Solution linearized }
	\end{equation}
	Integration constants are found from the BCs \cite{Kurpianov1988}
	\begin{equation}
	\partial_{z}\hat{f}_{\text{F},\pm}(\pm d_{\text{F}})=\pm 2\kappa_{b\text{F}}F_{\text{S}}\qty{\cos(\varphi/2)\hat{\tau}_{1}\pm\sin(\varphi/2)\hat{\tau}_{2}}
	\end{equation}	
	We find
	\begin{align}
	\hat{C}_{\pm}=&\frac{2\kappa_{b\text{F}}}{\kappa_{\text{F}\pm}}\frac{F_{\text{S}}\cos(\varphi/2)}{\sinh(\kappa_{\text{F}\pm}d_{\text{F}})}\hat{\tau}_{1}\\
	\hat{S}_{\pm}=&\frac{2\kappa_{b\text{F}}}{\kappa_{\text{F}\pm}}\frac{F_{\text{S}}\sin(\varphi/2)}{\cosh(\kappa_{\text{F}\pm}d_{\text{F}})}\hat{\tau}_{2}\\
	\end{align}
	One can write for \(\check{f}(\pm
	d_{\text{F}})\)
	\begin{equation}
	\check{f}(\pm
	d_{\text{F}})=2\kappa_{b\text{F}}F_{\text{S}}\qty(\cos(\varphi/2)\qty{\qty(\check{X}_{10}\Re\pm i\check{X}_{13}\Im)\frac{1}{\kappa_{\text{F},\omega}\tanh(\kappa_{\text{F},\omega}d_{\text{F}})}}+\sin(\varphi/2)\qty{\qty(\check{X}_{20}\Re\pm i \check{X}_{23}\Im)\frac{\tanh(\kappa_{\text{F},\omega}d_{\text{F}})}{\kappa_{\text{F},\omega}}})
	\end{equation}
	where \(\kappa_{\text{F},\omega}^{2}=2(\abs{\omega}+iJ_{\omega})/D_{\text{F}}\).\\
	The linearized Eq.(\ref{Appendix Strong PE: Correction S}) can be written as follows
	\begin{equation}
	-\partial_{zz}^{2}\delta g_{33}^{(\text{S})}+\kappa_{\text{S},\omega}^{2}\delta g_{33}^{(\text{S})}=0\label{Appendix Weak PE: Corection}
	\end{equation}
	The BC to Eq.(\ref{Appendix Weak PE: Corection}) is
	\begin{equation}
	\partial_{z} \delta g_{33}^{(S)}(\pm d_{\text{F}})=-\kappa_{b\text{S}}\qty(\check{g}_{\text{F}}-\check{G}_{\text{S}}\check{g}_{\text{F}}\check{G}_{\text{S}})_{33}
	\end{equation}
	or
	\begin{equation}
	\partial_{z} \delta g_{33}^{(S)}(\pm d_{\text{F}})=4\kappa_{b\text{S}}\kappa_{b\text{F}}G_{\text{S}}F_{\text{S}}^{2}i\Im{\frac{\cos[2](\varphi/2)+\sin[2](\varphi/2)\tanh[2](\kappa_{\text{F},\omega}d_{\text{F}})}{\kappa_{\text{F},\omega}\tanh[2](\kappa_{\text{F},\omega}d_{\text{F}})}}
	\end{equation}
	Here, we have taken into account that the correction \(g_{33}^{(\text{F})}\sim f_{13}^{2}\)
	is small. The solution for Eq.(\ref{Appendix Weak PE: Corection}) is
	\begin{equation}
	\delta g_{33}^{(\text{S})}(z)=\delta g_{33}^{(\text{S})}(\pm d_{\text{F}})\exp(-\kappa_{\text{F},\omega}\abs{z-d_{\text{F}}})
	\end{equation}
	with
	\begin{equation}
	\delta g_{33}^{(S)}(\pm d_{\text{F}})=4G_{\text{S}}F_{\text{S}}^{2}i\frac{\kappa_{b\text{S}}\kappa_{b\text{F}}}{\kappa_{\text{S},\omega}}\Im{\frac{\cos[2](\varphi/2)+\sin[2](\varphi/2)\tanh[2](\theta_{\text{F}})}{\kappa_{\text{F},\omega}\tanh[2](\theta_{\text{F}})}} \label{Appendix Weak PE: Solution correction}
	\end{equation}
	where we defined \(\theta_{\text{F}}=\kappa_{\text{F},\omega}d_{\text{F}}\).\\
	Using this expression we obtain Eq.(\ref{Eq.3:Weak M(phi)}) for the spin polarization in
	superconductors S.
	Close to $T_{c}$ the coefficient $\gamma _{\varphi }$ is equal to
	\begin{equation}
	\gamma_{\varphi}=4\sqrt{2}\frac{\xi_{\text{S}}}{d_{\text{F}}}\frac{\kappa_{b\text{S}}\kappa_{b\text{F}}}{\kappa_{\text{S},\omega}\kappa_{\text{F}c}}\frac{2T}{J}\Im{\frac{\cos[2](\varphi/2)+\sin[2](\varphi/2)\tanh[2](\theta_{\text{F}c})}{(1+i)\tanh(\theta_{\text{F}c})}}\frac{\Delta^{2}}{\pi T}\sum_{\omega\geq0}(2n+1)^{-3} \label{Appendix Weak PE: Gamma}
	\end{equation}
	where  \(\theta _{\text{F}c}=\kappa
	_{\text{F}c}d_{\text{F}}(1+i)/\sqrt{2}\), \(\kappa _{\text{F}c}=\sqrt{J/D_{\text{F}}}\), and \(
	\sum_{n}(2n+1)^{-3}=8\zeta(3)/8\) with the Riemann zeta function \(\zeta(x)\).
	
\setcounter{equation}{0}
\setcounter{figure}{0}
\setcounter{table}{0}
\makeatletter
\renewcommand{\theequation}{C\arabic{equation}}
\renewcommand{\thefigure}{C\arabic{figure}}
\renewcommand{\bibnumfmt}[1]{[C#1]}

	\section{Magnetostatics}
	
	The boundary conditions for Eq.(\ref{Eq.10: Boundary Condition Vectorpotential}) can be easily obtained from
	\begin{equation}
		\partial_{z} H_{\text{S}}(x,z)|_{z=\pm d_{\text{F}}}=\lambda_{\text{S}}^{-2}\qty[A_{\text{S}}(x,z)-\frac{\Phi_{0}}{2\pi}\partial_{x}\chi(x,z)]_{z=\pm d_{\text{S}}}
	\end{equation}
	Subtracting the expression for Eq.(\ref{Eq.7:currentphase})  at the interfaces \(z=d_{\text{F}}\) and \(z=-d_{\text{F}}\), we obtain Eq.(\ref{Eq.9: Boundary condition}) with \(\qty[A_{\text{S}}(x)]=A_{\text{S}}(x,d_{\text{F}})-A_{\text{S}}(x,-d_{\text{F}})\) and \(\varphi(x)=\chi(x,d_{\text{F}})-\chi(x,-d_{\text{F}})\). The difference \([A_{S}(x)]\) can be found by taking into
	account the variation of the vector potential in the F film. The latter can be
    easily found in the limit \(\lambda _{\text{F}}\gg d_{\text{F}}\) (see Ref.\cite{volkov_spin_2019})
	\begin{align}
		A_{\text{F}}(x,z)\cong&\qty(H_{0}(x)+4\pi M_{0})\qty(1+\frac{\kappa_{\text{F}}^{2}z^2}{2})z\\
		B_{\text{F}}(x,z)\cong&\qty(H_{0}(x)+4\pi M_{0})\qty(1+\frac{\kappa_{\text{F}}^{2}z^2}{2})\\
		H_{\text{F}}(x,z)\cong&H_{0}(x)\qty(1+\frac{\kappa_{\text{F}}^{2}z^2}{2})+4\pi M_{0}\frac{\kappa_{\text{F}}^{2}z^2}{2}\\
	\end{align}
	where we set \(M_{\text{F}}\approx M_{0}\)  since corrections to \(M_{0}\) in the F
	film due to the proximity effect  are small and do not significantly change the
	final results. For completness, we also write down the formulas for the fields \(B_{\text{F}}(x,z)\) and \(
	H_{\text{F}}(x,z)\).
	
\setcounter{equation}{0}
\setcounter{figure}{0}
\setcounter{table}{0}
\makeatletter
\renewcommand{\theequation}{D\arabic{equation}}
\renewcommand{\thefigure}{D\arabic{figure}}
\renewcommand{\bibnumfmt}[1]{[D#1]}

	\section{Critical Current}
	
	The josephson current density across the junction  can be be obtained using
	\begin{align}
		j_{J}(\varphi)=&i\sigma_{\text{F}}\frac{2\pi T}{e}\sum_{\omega\geq0}\frac{1}{4}\Tr{\check{X}_{30}\check{g}_{\text{F}}(z)\partial_{z}\check{g}_{\text{F}}(z)}\\
		=&i\sigma_{\text{F}}\frac{2\pi T}{e}\kappa_{b\text{F}}\sum_{\omega\geq0}\frac{1}{4}\Tr{\check{X}_{30}\qty[\check{g}_{\text{F}}(d_{\text{F}}),\check{g}_{\text{S}}]}
	\end{align}
	where \(\sigma_{\text{F}}\) is the conductivity of the F.\\
	In the second line, we used the boundary condition Eq.(\ref{Appendix Strong PE: Boundary Cond.}). At temperatures close to \(T_{c}\), the Josephson current \(
	j_{J}(\varphi )\) can be written  as follows
	\begin{equation}
		j_{J}(\varphi)=j_{c}\sin(\varphi)  \label{Appendix Critical Current: current}
	\end{equation}
	One can find the critical current density \(j_{c}\) in the limit of a strong and a weak PE
	by using Eqs.(\ref{Appendix Strong PE: Form GF in F},\ref{Appendix Weak PE: Solution linearized },)
	\begin{equation}
		j_{c}=\frac{\pi T}{eR_{b\square}}\sum_{\omega\geq 0}F_{\text{S}}^{2}\cong \frac{
		\pi \Delta ^{2}}{8eR_{b\square}T} \label{Appendic Critical Current: Strong PE}
	\end{equation}%
	where in the the case of a strong PE we get
	\begin{align}
		j_{c}=&\sigma_{\text{F}}\frac{2\pi T}{e}4\kappa_{b\text{F}}^{2}\Re{\sum_{\omega\geq 0}\frac{F_{\text{S}}}{\kappa_{\text{F},\omega}\sinh(2\theta_{\text{F}})}}\label{Appendix Critical Current: Weak PE}\\
		\cong&\frac{\pi \sigma_{\text{F}}}{\sqrt{2}}\frac{\Delta^2}{eT}\kappa_{b\text{F}}^{2}d_{\text{F}}\xi_{\text{F}}\frac{\sinh(\alpha)\cos(\alpha)-\cosh(\alpha)\sin(\alpha)}{\kappa_{\text{F}}\sinh(2\theta_{\text{F}})}
	\end{align}
	and in the case of a weak PE. Here \(\alpha=\sqrt{2}
	d_{\text{F}}/\xi_{\text{F}}\), \(\xi_{\text{F}}=\sqrt{D_{\text{F}}/J}\). For large \(\alpha\) we  can rewrite Eq.(\ref{Appendix Critical Current: Weak PE}) as
	\begin{align}
		j_{c}=&\sigma_{\text{F}}\frac{2\pi T}{e}4\kappa_{b\text{F}}^{2}\Re{\sum_{\omega\geq 0}\frac{F_{\text{S}}}{\kappa_{\text{F},\omega}\sinh(2\theta_{\text{F}})}}\\
		\cong&-\pi\sigma_{\text{F}}\frac{\Delta^{2}}{eT}\kappa_{b\text{F}}^{2}d_{\text{F}}\xi_{\text{F}}\exp(-\alpha)\sin(\alpha-\pi/4)
	\end{align}

		\subsection{Small Contribution of the Spin Polarization}
		
		We expand the phase difference \(\varphi(\tilde{x})\) up to the second order and write the integral in Eq.(\ref{Eq.23 :Current small}) in the form%
		\begin{align}
			I_{J}=&j_{c}\int_{0}^{1}\sin(\varphi(\tilde{x}))\dd{\tilde{x}}\\
			\cong&j_{c}\int_{1/2}^{-1/2}\qty[\sin(\varphi_{0})+\varphi_{1}\cos(\varphi_{0})-\varphi_{2}\sin(\varphi_{0})+\frac{\varphi_{1}}{2}\cos(\varphi_{0})]\dd{\tilde{x}}\label{Appendix Small Contribution: Current exp}
		\end{align}
		We need to find \(\varphi_{i}\) and \(\tilde{\Phi}_{m}\). In the zeroth order approximation  we obtain
		\begin{equation}
		\varphi _{0}=\tilde{\Phi}_{m,0}\tilde{x}+c \label{Appendix Small Contribution: zeroth order approx}
		\end{equation}%
		where \(\tilde{\Phi}_{m,0}=2\pi(H_{ext}(2d_{\text{F}}+2\lambda_{\text{S}})+4\pi M_{0} 2d_{\text{F}})/\Phi_{0}\). In the first approximation, we get from Eqs.(\ref{Eq.18: DGL FLUX},\ref{Eq.19: DGL PHI})
		\begin{equation}
			\tilde{\Phi}_{m,1}=-\frac{r}{\tilde{\Phi}_{m,0}}\cos(\varphi_{0}) \qquad \varphi_{1}=-\frac{\tilde{p}}{\tilde{\Phi}_{m,0}}\sin(\varphi_{0})\label{Appendix Small Contribution: first order approx}
		\end{equation}
		where \(\tilde{p}=p+r/\tilde{\Phi}_{m,0}\). For the second order contribution we obtain
		\begin{align}
			\tilde{\Phi}_{m,2}=&\frac{r\tilde{p}}{(2\tilde{\Phi}_{m,0})^{2}}\cos(2\varphi_{0})\\
			\varphi_{2}=&-\frac{p\tilde{p}}{2\tilde{\Phi}_{m,0}}\tilde{x}+\frac{\tilde{p}}{(2\tilde{\Phi}_{m,0})^{2}}\qty(p+\frac{r}{2\tilde{\Phi}_{m,0}})\sin(2\varphi_{0})\label{Appendix Small Contribution: second order approx}
		\end{align}
		The current \(\tilde{I}_{J}\equiv I_{J}/j\) can be written as
		\begin{equation}
			\tilde{I}_{J}=\tilde{I}_{Fr}+\delta\tilde{I}_{0}+\delta\tilde{I}_{2}+\delta\tilde{I}_{3}\label{Appendix Small Contribution: Current total}
		\end{equation}
		where 
		\begin{align}
			\tilde{I}_{Fr}=&\int_{1/2}^{-1/2}\sin(\varphi_{0})\dd{\tilde{x}}\\
			\delta I_{0}=&\delta a \sin(c)\\
			\delta
			I_{2}=&\int_{1/2}^{-1/2}\qty[\varphi_{2}\cos(\varphi_{0})-\frac{\varphi_{1}^{2}}{2}\sin(\varphi_{0})]\dd{\tilde{x}}\label{Appendix Small Contribution: Current corrections}\
		\end{align}
		By using Eqs.(\ref{Appendix Small Contribution: Current exp}-\ref{Appendix Small Contribution: Current corrections}), we find
		\begin{align}
			\tilde{I}_{0}=&a_{Fr}\sin(c)\\
			\delta\tilde{I}_{0}=&\delta a \sin(c)\\
			\delta\tilde{I}_{2}=&-\lambda_{2}\sin(2c)\\
			\delta\tilde{I}_{3}=&\delta a\tilde{I}_{0}+\lambda_{3}\sin(3c)
		\end{align}	
		where \(a=a_{Fr}+\delta a\), \(a_{Fr}=(2/\tilde{\Phi}_{m,0})\sin(\tilde{\Phi}_{m,0}/2)\), \(\delta a = \qty{-\tilde{p}^{2}/2\tilde{\Phi}_{m,0}^{3}+\tilde{p}r/(2\tilde{\Phi}_{m,0})^{4}}\sin(\tilde{\Phi}_{m,0}/2)\), \(\lambda_{2}=-2\tilde{p}/(2\tilde{\Phi}_{m,0})^{2}\sin(\tilde{\Phi}_{m,0})\) and \(\lambda_{3}=\qty{(2p+3r/2\tilde{\Phi}_{m,0})\tilde{p}/(12\tilde{\Phi}_{m,0}^{3})}\sin(3\tilde{\Phi}_{m,0}/2)\). The quantities \(\lambda\) are assumed to be small: \(\lambda<<1\). We have to find the maximum of \(\tilde{I}_{J}\) as a function of the constant \(c\) by expanding the current in powers of \(\lambda\): \(c=c_{0}+c_{1}+c_{2}\). Calculating the derivative \(\partial\tilde{I}_{J}/\partial c\), we find
		\begin{equation}
			c_{0}=\frac{\pi}{2},\quad c_{1}=-2\frac{\lambda_{1}}{a}, \qquad c_{2}=0
		\end{equation}
		Thus the maximal current is equal to
		\begin{equation}
			\tilde{I}_{J}=\frac{2}{\tilde{\Phi}_{m,0}}\sin(\tilde{\Phi}_{m,0}/2)+\delta \tilde{I}_{J}
		\end{equation}
		The first term is the standard Fraunhofer pattern and the second term is a
		correction due to spin polarization (\(\sim p\)) and due to the finite length \(L\)
		compared to the Josephson length \(\lambda _{J}\). This correction is
		\begin{equation}
			\delta \tilde{I}_{J}=\sin(\tilde{\Phi}_{m,0}/2)\qty[-\frac{2\tilde{p}}{(\tilde{\Phi}_{m,0})^{3}}+\frac{pr}{(2\tilde{\Phi}_{m,0})^{4}}]+\frac{\tilde{p}}{2\tilde{\Phi}_{m,0}^{3}}\sin(\tilde{\Phi}_{m,0})\cos(\tilde{\Phi}_{m,0}/2)-\frac{\tilde{p}(\tilde{p}+(p+r/2\tilde{\Phi}_{m,0}))}{24\tilde{\Phi}_{m,0}^{3}}\sin(3\tilde{\Phi}_{m,0}/2)
		\end{equation}

\end{document}